\documentclass[journal]{IEEEtran}
\usepackage{cite}
\usepackage{graphicx}
\usepackage{amsmath}
\usepackage{amssymb}
\usepackage{algorithm}

\begin{document}

\title{A Channel Coding Approach for Physical-Layer Authentication}

\author{Xiaofu Wu\IEEEauthorrefmark{1} and \IEEEauthorblockN{Zhen Yang\IEEEauthorrefmark{1}} \\
\IEEEauthorblockA{\IEEEauthorrefmark{1}
Nanjing University of Posts and Telecommunications, Nanjing 210003, CHINA\\ Emails: xfuwu@ieee.org, yangz@njupt.edu.cn}}

%\author{Xiaofu Wu\IEEEauthorrefmark{1}, \IEEEauthorblockN{Zhen Yang\IEEEauthorrefmark{1} and
%Lu Gan\IEEEauthorrefmark{2}} \\
%\IEEEauthorblockA{\IEEEauthorrefmark{1}
%Nanjing University of Posts and Telecommunications, Nanjing 210003, CHINA\\ Email: xfuwu@ieee.org, and yangz@njupt.edu.cn}
%\\ \IEEEauthorblockA{\IEEEauthorrefmark{2}Brunel University, London UB8 3PH, UK\\ Email: lu.gan@brunel.ac.uk}}
%
%\author{Xiaofu~Wu and Zhen~Yang % <-this % stops a space
%\thanks{This  work was supported in part by the National Science Foundation of China under Grants 61372123, 61271335, 61032004. The work of Wu was also supported by the Scientific Research Foundation of Nanjing University of Posts and Telecommunications under Grant NY213002.}% <-this % stops a space
%\thanks{Xiaofu~Wu and Zhen~Yang are with the Institute of Signal Processing and Transmission, Nanjing University of Posts and Telecommunications, Nanjing 210003, China (Emails:
%        xfuwu@ieee.org, yangz@njupt.edu.cn)).}}
%\thanks{Lu~Gan is with the School of Engineering and Design, Brunel University, London UB8 3PH, UK (e-mail: lu.gan@brunel.ac.uk).}}

%\markboth{IEEE ITW'2006}{Shell \MakeLowercase{\textit{et al.}}:
%Bare Demo of IEEEtran.cls for Journals}

\maketitle

\begin{abstract}
For physical-layer authentication, the authentication tags are often sent concurrently with messages without much bandwidth expansion. In this paper, we present a channel coding approach for physical-layer authentication.  The generation of authentication tags can be formulated as an encoding process for an ensemble of codes, where the shared key between Alice and Bob is considered as the input and the message is used to specify a code from the ensemble of codes.  Then, we show that the security of physical-layer authentication schemes can be analyzed through decoding and physical-layer authentication schemes can potentially achieve both information-theoretic and computational securities.
\end{abstract}


\IEEEpeerreviewmaketitle

\section{Introduction}

\PARstart{M}{essage} authentication codes (MACs) are cryptographic primitives used extensively in the construction of
security services, include authentication, nonrepudiation, and integrity. Basically, message authentication is to ensure that an accepted message
truly comes from its acclaimed transmitter. When the transmitter intends to send a message, it also generates a MAC, which is a function of the message and a shared key, known only to both the transmitter and the receiver. The generated MAC is often appended to the message \cite{CyptBook}.  At the receiver, a MAC is computed from the received message and compared to the MAC that is transmitted. If the two MACs are identical, then the transmitter is identified as a legal user and it is highly likely the received message is exactly equal to the one transmitted.

MAC algorithms can be constructed from other cryptographic primitives, such as cryptographic hash functions or from block cipher algorithms. Currently, the security of MAC algorithms rely on the hardness of hush functions, i.e, given the message and its MAC, it is ``hard" to forge a MAC on a new message.

As the development of mobile communications, ensuring security of wireless communications has becoming increasingly important. Openness of wireless networks makes them vulnerable to spoofing attacks where an unauthorized user masquerades as another legitimate user. Although conventional cryptographic security mechanisms can be used to foil such attacks above the physical layer \cite{ComMagAttack,ProcIEEEAttack}. However, it was believed that more efforts should be done to prevent potential innovative attacks since the wireless medium offers novel avenues for intrusion. In recent years, there has been various efforts \cite{XiaoPHY,DanPHY,YuIFS,TugnaitJSAC} in authenticating the sender and receiver at the physical layer, based on prior coordination or secret sharing, where the sender is authenticated if the receiver can successfully demodulate and decode the transmission. Among various reported works, it was commonly observed that the physical properties of the wireless medium are a powerful source of domain-specific information that can be used to complement and enhance traditional security mechanisms\cite{XiaoPHY}.

In \cite{YuIFS}, a physical-layer authentication scheme was proposed, in which MACs, along with messages, are transmitted concurrently over the physical layer. Compared to the traditional transmission approach above the physical layer, the authors claims the possibility of information-security due to the introduction of the noise. However, this is not justified in a rigourous way.

In this paper, we provide a channel coding approach for physical-layer authentication. Our contributions include two aspects. Firstly, the computational security can be expressed as the requirement for decoding complexity. Secondly, the information security can be formulated for physical-layer authentication schemes using the standard techniques for a converse proof of channel coding theorem \cite{ITCoverBook}.

Throughout this paper, upper case letters (e.g., $X$ ) will denote random variables, lower case letters (e.g., $x$) will denote
realizations of the corresponding random variables. and calligraphic letters (e.g., $\mathcal{X}$) will denote finite alphabet sets over
which corresponding variables range.  Also, upper case boldface letters (e.g., $\mathbf{X}$ ) will denote random vectors whereas lower case
boldface letters (e.g., $\mathbf{x}$) will denote realizations of the corresponding random vectors.

\section{A Coding Formulation of Physical-Layer Authentication}
\subsection{Physical-Layer Authentication}
%\begin{figure*} %[htbp]
%   \centering
%   \includegraphics[width=0.82\textwidth]{PhyAuOFDM.eps}
%   \caption{Channel-Phase Based Challenge-Response Authentication for OFDM Transmission.}
%   \label{fig:codingView}
%\end{figure*}
%\begin{figure}[htb] %[htbp]
%   \centering
%   \includegraphics[width=0.5\textwidth]{zetaPdf20dB.eps} %{PWBF_BER_Uth.eps}
%   \caption{$\bar{\rho}$ versus signal-to-noise ratio with Rayleigh fading SNR=20dB.}
%   \label{fig:num2}
%\end{figure}

Suppose that Alice and Bob agree on a keyed authentication scheme that allows Bob
to verify that the messages he receives are from Alice. In order to authenticate, Alice sends an authentication tag (or a MAC), along with a message, for declaring his identity. We call the transmitted signal under this scheme as the tagged signal.

Formally, Alice, as the sender, wants to transmit the authentication tag $T$ together with the message $S$ so Bob (as the receiver) can
verify her identity. In general, the tag is a function of the
message $S$ and the secret key $K$
\begin{eqnarray}
  \label{eq:cn}
     T= \tau(S, K),
\end{eqnarray}
where $\tau: \mathcal{S} \times \mathcal{K} \rightarrow \mathcal{T}$ is a (hash) function.

In order to focus on the essential ideas, we assume here that both the message $S$ and tag $T$ can be denoted as binary random vectors with BPSK modulation.
In what follows, we do not discriminate between binary and bipolar vectors, as it can be well understood from the text.

Often, the tag is a short string computed on the message $S$ to be authenticated and the shared secret key $K$. Let $L_s, L_k, L_t$  denote the length of the message $S$, the key $K$ and the tag $T$, respectively. In practice, $L_s \gg L_t$. Here, we always assume that $L_s = Q L_t$, where $Q$ is a large integer.

The tag is padded to the message and simultaneously
transmitted. The tagged signal (in a discrete column vector form) can be written as
\begin{eqnarray}
  \label{eq:cn}
     \mathbf{u} &=& \rho_s \mathbf{s} + \rho_t \mathbf{t}_q,
\end{eqnarray}
where $0 < \rho_s, \rho_t < 1$, $\rho_s^2 + \rho_t^2 =1$, and $\mathbf{t}_q = \psi(\mathbf{t})$ is a modulation process for modulating a binary tag string $\mathbf{t}$ into the physical discrete-signal vector $\mathbf{t}_g$, which is chosen to meet $E[\mathbf{s}^H \mathbf{t}_q]=0$ \cite{YuIFS}.  Hence,we can interpret $\rho_s^2$ and $\rho_t^2$ as energy allocations of the message and tag, respectively.

Essentially, for concurrent transmission of both message and tag,  the bit string of a tag should be transmitted with a much lower rate ($\frac{1}{Q}$) than the message symbol rate. In \cite{YuIFS}, the Haar wavelet is employed for modulating tags. In this paper, we, however, assume a simple repetition function of $\psi(\cdot)$, namely, each component of $\mathbf{t}$ is repeated $Q$ times, which means that
\begin{eqnarray}
  \label{eq:cn}
     \psi(\mathbf{t}) = [\underbrace{t_1,\cdots,t_1}, \underbrace{t_2,\cdots,t_2},\cdots,\underbrace{ t_{L_k},\cdots,t_{L_k}}]^T.
\end{eqnarray}
This is employed for ease of analysis.

By assuming an additive white Gaussian noise (AWGN) channel model, the received signal vector at Bob can be written as
\begin{eqnarray}
  \label{eq:cn}
     \mathbf{r} &=& \mathbf{u} + \mathbf{z},
\end{eqnarray}
where $\mathbf{z}$ is assumed to an AWGN vector.

As $|\frac{\rho_s}{\rho_t}| \gg 1$ and the signal-to-noise ratio (SNR) is sufficiently high,  the transmitted message is assumed to be completely recoverable (often enhanced by error-correcting codes) for both Bob and Eve. Therefore, one can assume that the transmitted signal vector $\mathbf{s}$ is known to both Bob and Eve.

When $\mathbf{s}$ is available at the receiver, it can cancel the message from the received signal samples and the message-free version of a tag can be retrieved \cite{YuIFS}, which takes the form of
\begin{eqnarray}
  \label{eq:cn}
     \mathbf{y} = \mathbf{x} + \mathbf{w},
\end{eqnarray}
where $\mathbf{x}=\mathbf{t}$ and  $\mathbf{w}$ is the zero-mean additive white Gaussian noise vector with variance
$E[\mathbf{w}_i^\dag \mathbf{w}_j]=\delta_{ij} \gamma_t^{-1} I_{L_t}$ and $\gamma_t$ denotes the signal-to-noise ratio (SNR) observed by the authentication tags.

\subsection{A Coding Formulation}
Given a message $\mathbf{s} \in \mathcal{S}$, it is possible to generate a code $\mathcal{C}(\mathbf{s})$, which comprised of $2^{L_k}$ codewords, namely,
\begin{eqnarray}
\mathcal{C}(\mathbf{s})=\{\mathbf{c}_1(\mathbf{s}), \cdots, \mathbf{c}_{2^{L_k}}(\mathbf{s})\},
\end{eqnarray}
where each codeword $\mathbf{c}_k(\mathbf{s})=\tau\left(\mathbf{s},\mathbf{k}\right)$ is indexed by a possible key $\mathbf{k}\in \mathcal{K}$ with $k-1=\kappa(\mathbf{k})$ denoting the decimal number expression of the binary vector $\mathbf{k}$. There are $|\mathcal{K}|=2^{L_k}$ codewords.
Now, the code rate of $\mathcal{C}(\mathbf{s})$ can be defined as
\begin{equation}
  R_c = \frac{L_k}{L_t}.
\end{equation}

 Consider that Alice wants to authenticate with Bob,  she normally sends a message $\mathbf{s}$, and then a tag $\mathbf{c}_k(\mathbf{s})=\tau\left(\mathbf{s},\mathbf{k}\right)$ is generated using the shared key $\mathbf{k}$. Equivalently, the generation of the tag for a given message $\mathbf{s}$ can be considered as an encoding process of
\begin{eqnarray}
  \tau(\mathbf{s},\cdot): \mathcal{K} \rightarrow \mathcal{T}.
\end{eqnarray}
As the message $\mathbf{s}$ is generated according to a finite message set $\mathcal{S}$, one have to consider an ensemble of codes $\Omega(\mathcal{C})=\{\mathcal{C}(\mathbf{s}): \mathbf{s} \in \mathcal{S}\}$, which is of fixed rate $R_c$.

This ensemble of codes $\Omega(\mathcal{C})$ is revealed to both Alice and Bob. From a standard cryptographic view, this code ensemble is also revealed to Eve.

\newtheorem{defn}{Definition}
\begin{defn}
The minimum Hamming distance of the code ensemble $\Omega(\mathcal{C})$ can be defined as
\begin{eqnarray}
  \label{eq:cn}
     d_{\min}\left(\Omega(\mathcal{C})\right) = \min_{\mathbf{s} \in \mathcal{S}}\min_{\mathbf{k} \neq \hat{\mathbf{k}}} d_H\left(\tau(\mathbf{s},\mathbf{k}),\tau(\mathbf{s},\hat{\mathbf{k}})\right),
\end{eqnarray}
where $d_H(\mathbf{c}_1,\mathbf{c}_2)$ denotes the Hamming distance of two binary vectors $\mathbf{c}_1$ and $\mathbf{c}_2$.
\end{defn}

Now, the task of physical-layer authentication can be formally formulated as a hypothesis testing problem as follows.

{\ }
\begin{enumerate}
\item[$\clubsuit$]
     Bob decides if $\mathbf{y}$ is from Alice or not by assuming that  \\ $\mathbf{s}, \mathbf{k}, \tau(\cdot,\cdot)$ are available;
\item[$\spadesuit$]
     Eve tries to retrieve $\mathbf{k}$ from $\mathbf{y}$ by assuming that \\ $\mathbf{s}, \tau(\cdot,\cdot)$ are available.
\end{enumerate}

\newtheorem{lem}{Lemma}

\section{Hypothesis Testing}
\subsection{General Formulation}
 To complete the authentication process, Bob requires to verify that whether the response signal $\mathbf{y}$ is from Alice or not. If the response signal is not from Alice but Eve (an impersonation attacker), it is assumed that Eve generates length-$L_k$ binary random vector $\mathbf{k}_E$ for authentication as there is no any information about $\mathbf{k}_A (= \mathbf{k}_B)$ available. Essentially, this is cast as a binary hypothesis testing problem:
\begin{eqnarray}
  \label{eq:hp}
     H_0 &:&  K = \mathbf{k}_B \\
     H_1 &:&  K = \mathbf{k}_E
\end{eqnarray}
where $K$ denotes the acknowledged key.

Hypothesis testing is the task of deciding which of two hypotheses,
$H_1$ or $H_0$, is true, when one is given the value $u$ of a random variable $U$ (e.g., the outcome of a measurement), namely, $U=u$.
In our case, $U=(Y, K)$, $u=(\mathbf{y},\mathbf{k}_B)$. We begin with the formulation of the optimum binary hypothesis testing, i.e.,
\begin{eqnarray}
\label{eq:opmHyp}
  \eta &=&\log\frac{p_{H_0}(U=u)}{p_{H_1}(U=u)} = \log\frac{p(\mathbf{y}, K=\mathbf{k}_B)}{p(\mathbf{y}) p(K=\mathbf{k}_B)} \\
      &=&\log\frac{p(\mathbf{y}|K=\mathbf{k}_B)}{\sum_{\mathbf{k} \in \mathcal{F}_2^{L_k}} p(\mathbf{y}|K=\mathbf{k}) p(K=\mathbf{k})}.
\end{eqnarray}
We point out that in the case of $H_1$, the generation of the message and key is independent to each other as there is no means to efficiently gauss the key.

As the message $\mathbf{s}$ is assumed to be available, it is clear that
\begin{eqnarray}
  p(\mathbf{y}|\mathbf{k}) \propto \exp\left[-\frac{(\mathbf{y}-\mathbf{t})^\dag (\mathbf{y}-\mathbf{t})}{2\sigma_w^2}\right]
\end{eqnarray}
with $\mathbf{t}=\tau(\mathbf{s},\mathbf{k})$.

In general, this binary hypothesis testing problem in its optimum form can not be easily tackled as it requires to enumerate $2^K$ binary vectors of $\mathbf{k}$ with a priori uniform distribution. However, its performance can be information-theoretically  bounded \cite{Maurer}, which is summarized as follows.

\subsection{Detection Probability vs. False Alarm Probability}

Let $P_D=1-\alpha$ be the detection probability, namely, the probability of successful declaration of $H_0$ when $H_0$ is actually true, and $P_f= \beta$ be the false alarm probability, namely, the probability of false declaration of $H_0$ when $H_1$ is actually true.

Let the function $d(\alpha, \beta)$ be defined by
\begin{eqnarray}
  \label{eq:cn}
     d(\alpha, \beta) = \alpha \log \frac{\alpha}{1-\beta} +  (1-\alpha) \log \frac{1-\alpha}{\beta}.
\end{eqnarray}

With optimal hypothesis testing (\ref{eq:opmHyp}), its detection probability and false alarm probability are closely connected.
%\newline

%\newtheorem{lem}{Lemma}
\begin{lem}\cite{Maurer}
The detection probability $1-\alpha$ and the false alarm probability $\beta$ satisfy
\begin{eqnarray}
  \label{eq:cn}
     d(\alpha,\beta) \le D_{KL}\left( p(\mathbf{y},\mathbf{k}_B)||p(\mathbf{y})p(\mathbf{k}_B)  \right) = I\left(Y;K\right)
\end{eqnarray}
where $ I\left(Y;K\right)$ denotes the mutual information between two random variables $Y$ and $K$, and
\begin{eqnarray}
  \label{eq:cn}
      D_{KL}\left( f(x)||g(x) \right) = \sum_x f(x) \log \frac{f(x)}{g(x)}
\end{eqnarray}
for two probability distributions $f(x),g(x)$.
\end{lem}

\subsection{A Suboptimal Solution}
As the optimum hypothesis testing is difficult to implement, we propose to use a simple test statistic
\begin{eqnarray}
  \label{eq:cn}
   \eta = \mathbf{c}_B^T \mathbf{y},
\end{eqnarray}
and $\zeta$ is further compared to a threshold value $\varrho$ for making a final decision, where $\mathbf{c}_B = \tau(\mathbf{s},\mathbf{k}_B)$ is the codeword due to the input of $\mathbf{k}_B$ at Bob.

This approach can be viewed as a code acquisition approach encountered in code-division multiple-access (CDMA) communication systems, where $\mathbf{c}_B$ can be considered as a unique PN code, which is available at the sides of both Alice and Bob, but keeps unknown to any potential attacker.

In both hypotheses, $\eta$ is the sum of $L_t$ normally distributed random variables, which is still normally distributed. Therefore, it suffices to compute its mean and variance.

In the case of hypothesis $H_0$, one can show that
\begin{eqnarray}
  \label{eq:cn}
     \eta|H_0 = L_t + z_0,
\end{eqnarray}
where $z_0 = \sum_{i=1}^{L_t} c^B_i w_i$. We denote its mean and variance as
\begin{eqnarray}
  \label{eq:cn}
     \bar{\eta}_0 &\triangleq& E\{\eta|H_0\} = L_t, \nonumber \\
     \sigma^2_{H_0}&\triangleq&  \text{Var}\{\eta|H_0\} = L_t \gamma_t^{-1}.
\end{eqnarray}

By decomposing the hypothesis $H_1$ into a series of sub-hypothesises $\left\{H_1^{sk}: H_1, S=\mathbf{s}, K=\mathbf{k}_E\right\}$, i.e.,  by further assuming that the transmitted signal is $\mathbf{s}$ and Eve impersonates Alice using the key $\mathbf{k}_E$,  we have
\begin{eqnarray}
  \label{eq:etaH1}
     \eta|H_1^{sk} = L_t- 2 d_H\left(\tau(\mathbf{s},\mathbf{k}_B),\tau(\mathbf{s},\mathbf{k}_E)\right) + z_1,
\end{eqnarray}
where $z_1 = \sum_{i=1}^{L_t} c^B_i w_i$. Then,
\setlength{\arraycolsep}{0.0em}
\begin{eqnarray}
  \label{eq:cn}
     \bar{\eta}_1^{sk} &\triangleq& E\{\eta|H_1,\mathbf{s},\mathbf{k}_E\} = L_t-2 d_H\left(\tau(\mathbf{s},\mathbf{k}_B),\tau(\mathbf{s},\mathbf{k}_E)\right), \nonumber \\
     \sigma^2_{H_1^{sk}}&\triangleq& \text{Var}\{\eta|H_1,\mathbf{s},\mathbf{k}_E\} = L_t \gamma_t^{-1}.
\end{eqnarray}
\setlength{\arraycolsep}{5pt}
It is clear that $\eta|H_0 \sim \mathcal{N}\left(\bar{\eta}_0,\sigma^2_{H_0}\right)$ and  $\eta|H_1^{sk} \sim \mathcal{N}\left(\bar{\eta}_1^{sk},\sigma^2_{H_1^{sk}}\right)$.

The authentication is typically claimed if $\eta\ge \varrho$. The threshold $\varrho$ of this test is determined for a false alarm
probability $\beta$ according to the distribution of $\eta|H_1$
\begin{eqnarray}
    \varrho = \arg \min_{\varrho'}E_{\mathbf{s},\mathbf{k}_E}\left[Q\left(\frac{ \varrho'-\bar{\eta}_1^{sk}}{\sigma_{H_1^{sk}}}\right)\right] \ge  \beta,
\end{eqnarray}
where
\begin{equation}
 \label{eq:8}
 Q(x)=\frac{1}{\sqrt{2\pi}}\int_x^{\infty}
 \exp\left(-\frac{t^2}{2}\right)dt.
\end{equation}

The detection probability can be simply computed as
\begin{eqnarray}
        P_D = Q\left(\frac{\varrho - \bar{\eta}_0}{\sigma_{H_0}}\right).
\end{eqnarray}

\section{A Decoding Approach for Security Analysis}
For a physical-layer authentication system, we can characterize it using a quadruple $\left\{\mathcal{S},\mathcal{K},\Omega(\mathcal{C}), p(\mathbf{y}|\mathbf{x})\right\}$. In this paper, we always assume a memory-less channel and hence, $p(\mathbf{y}|\mathbf{x})=\prod_{i=1}^{L_t} p(y_i|x_i)$.

\subsection{Adversary Model}
Eve, as the adversary, is an aware receiver and knows the authentication scheme that Alice and Bob are
using. However, she does not know the shared secret key between Alice and Bob. She  can be a passive attacker or active attacker.
As an active attacker, Eve can perform impersonation attacks.

\subsection{Passive Attacks: A Decoding Approach for Recovery of Key}

As a passive attacker, Eve only monitors all frames inside the network during authentication, and tries to learn $\mathbf{k}_B$ from whatever it gets.

Firstly, we consider the noiseless setting as in a classic authentication application above the physical layer, in which Eve can directly acquire the signal $\mathbf{s}$ and the tag
$\mathbf{y}=\tau(\mathbf{s}, \mathbf{k}_B)$.

Given $\mathbf{s}$ and if the encoding rule
\begin{eqnarray*}
  \tau(\mathbf{s},\cdot): \mathcal{K} \rightarrow \mathcal{T}
\end{eqnarray*}
is a bijection, Eve can recover the key $\mathbf{k}$ by generating a lookup table of size $2^{L_k}$ and searching over this table for finding the key $\mathbf{k}_E$, which admits $\mathbf{y}=\tau(\mathbf{s}, \mathbf{k}_E)$.

In the language of coding, it means that the recovery of key can be considered as decoding of the received signal $Y$ to its maximum possible encoding input $\hat{K}(Y)$. Given $\mathbf{s}$, if any decoder $\hat{K}(Y)$ is of computational complexity $\mathcal{O}(2^{L_k})$, we claim that the computational security can be achieved for this authentication system.

\begin{defn} (Computational security) Given a  physical-layer authentication system $\left\{\mathcal{S},\mathcal{K},\Omega(\mathcal{C}), p(\mathbf{y}|\mathbf{x})\right\}$, we claim that this system is computationally secure if for any decoder $\hat{K}(Y)$, its computation complexity is of $\mathcal{O}(2^{L_k})$.
\end{defn}
{\ }

For ensuring computational security, it requires that  no any efficient decoding algorithm exists for any code $\mathcal{C}(\mathbf{s}) \in \Omega(\mathcal{C})$. Since the publication of Shannon's original paper in 1948, the search of the codes for achieving the channel capacity has come a long way. Currently, linear codes and their efficient decoding algorithms have been extensively studied. Therefore, for construction of a good physical layer authentication system, linear code ensembles should be better avoided as their complexity can often be reduced due to the linearity of codes.

In the classic authentication scenarios, Eve can observe several pairs of (message,tag), namely, $(\mathbf{s}_i, \mathbf{t}_i =\tau(\mathbf{s}_i,\mathbf{k})), i=1,\cdots,I$. For computational security, it means that Eve is still hopeless for getting an estimate of $\mathbf{k}$ with many observation pairs $(\mathbf{s}_i, \mathbf{t}_i)$. In the language of coding, this can be well justified as each pair $(\mathbf{s}_i, \mathbf{t}_i)$ reflects an codeword of $\mathcal{C}(\mathbf{s}_i)$. If $\mathbf{s}_i \neq \mathbf{s}_j$, $\mathbf{t}_i$ and $\mathbf{t}_j$ reveal the structure of two different codes, i.e., $\mathcal{C}(\mathbf{s}_i)$ and $\mathcal{C}(\mathbf{s}_j)$.

{\ }

Secondly, we consider the noise setting, as seen in the physical-layer authentication scenarios.
%\newline
%\newtheorem{defn}{Definition}
\begin{defn}
Let the binary codeword $\mathbf{c}\in C$, which is further modulated with $\mathbf{x}(\mathbf{c})$ and transmitted over the channel $p(\mathbf{y}|\mathbf{x})$, the received vector $\mathbf{y} \in \mathcal{R}^{L_t}$. A maximum-likelihood (ML) decoding algorithm decodes
the vector $\mathbf{y}$ into a codeword $\hat{\mathbf{c}}$, such that
\begin{eqnarray}
  \label{eq:cn}
     \hat{\mathbf{c}} = \max_{\mathbf{c} \in \mathcal{C}} p\left(\mathbf{y}|\mathbf{x}(\mathbf{c})\right).
\end{eqnarray}
\end{defn}

\begin{defn} (ML recoverable) Given $\mathbf{y} \in \mathcal{R}^{L_t}$ and $\mathbf{s}$, where $\mathbf{y}=\tau(\mathbf{s}, \mathbf{k}) + \mathbf{w}$. For an ML decoder $\hat{\mathbf{k}}(\mathbf{y})$,  we mean that
\begin{eqnarray}
  \label{eq:cn}
     \hat{\mathbf{k}} = \max_{\mathbf{k} \in \mathcal{K}} p(\mathbf{y}|\mathbf{k},\mathbf{s}).
\end{eqnarray}
If $\Pr(\hat{\mathbf{k}} \neq \mathbf{k}_A)=0$, we claim that the authentication key is ML recoverable.
\end{defn}
{\ }

In what follows, we consider a binary-input continuous-output AWGN channel (Bi-AWGN). Its capacity $C_2\left(\gamma_t\right)$ is a function of the signal-to-noise ratio $\gamma_t$, which can be explicitly expressed as
\begin{eqnarray*}
       C_2 (\gamma_t) = \left[1 - \frac{1}{\sqrt{2\pi}} \int_{-\infty}^{\infty} e^{-(y-\beta)^2/2} \log_2 \left(1 + e^{-2\beta y}\right)dy\right],
\end{eqnarray*}
where $\beta=\sqrt{2 \gamma_t}$.

The SP59 bound of Shannon \cite{SPB1959} provides a lower bound on the decoding error probability of block codes transmitted over the AWGN channel. With a coding approach for physical-layer authentication, the best possible decoding probability with ML decoding for a potential eavesdropper can now be lower bounded with the Shannon's 1959 sphere-packing bound.

\begin{lem} (The 1959 Sphere-Packing Lower Bound \cite{SPB1959}) For a physical-layer authentication system, characterized by the quadruple $\left\{\mathcal{S},\mathcal{K},\Omega(\mathcal{C}), p(\mathbf{y}|\mathbf{x})\right\}$. Let a message $\mathbf{s}\in \mathcal{S}$ be sent, and the authentication tags are assumed to be transmited over a Bi-AWGN channel with the signal-to-noise ratio of $\gamma_t$. For any decoder $\hat{K}$, it is clear that $K\rightarrow \tau(\mathbf{s},K) \rightarrow X \rightarrow Y \rightarrow \hat{K}$ form a Markov process. Let $P_e = \Pr(K \neq \hat{K})$, we have that
\begin{eqnarray*}
  \label{eq:cn}
   P_e > P_{SPB}\left(L_t,\theta,\gamma_t\right),
\end{eqnarray*}
where
 \begin{eqnarray*}
      P_{SPB}\left(L_t,\theta,\gamma_t\right) = Q(\sqrt{2L_t\gamma_t}) + \frac{L_t-1}{\sqrt{2\pi}}e^{-L_t \gamma_t} \nonumber \\
         \cdot \int_{\theta}^{\pi/2}\sin(\phi)^{L_t-2} f_{L_t}(\sqrt{2L_t\gamma_t}\cos(\phi))d\phi,
\end{eqnarray*}
\begin{equation*}
   f_L(x)=\frac{1}{2^{\frac{L-1}{2}} \Gamma\left(\frac{L+1}{2}\right)} \int_0^\infty z^{L-1}\exp\left(-\frac{z^2}{2}+zx\right)dz,
\end{equation*}
and $\theta \in [0,\pi]$ satisfies the inequality  $2^{-L_t R} \le \frac{\Omega_{L_t}(\theta)}{\Omega_{L_t}(\pi)}$ with
\begin{eqnarray*}
   \Omega_{L_t}(\theta) = \frac{2\pi^{\frac{{L_t}-1}{2}}}{\Gamma(\frac{L_t-1}{2})} \int_0^\theta (\sin(\phi))^{L_t-2} d\phi.
\end{eqnarray*}

\end{lem}
{\ }

The SP59 bound is exponentially increased with the block length and the exponent is strictly negative for all $R_c > C_2(\gamma_t)$, it become clear that above capacity the minimum probability of error goes to 1 exponentially fast with the block length. Hence, one can achieve the information security for physical-layer authentication, which, however, not the case for classic authentication.

\begin{lem} (Information security)  Given a  physical-layer authentication system $\left\{\mathcal{S},\mathcal{K},\Omega(\mathcal{C})\right\}$ over an AWGN channel of the SNR $\gamma_t$, we claim that this system can achieve information security if $R_c > C(\gamma_t)$ when $L_t \rightarrow \infty$.
\end{lem}
{\ }

Numerically, we'll show that the decoding error probability can go to 1 even with short block length if the signal-to-noise ratio $\gamma_t$ is sufficiently low.

\subsection{Impersonation Attacks}
In a so-called impersonation attack at time $i$ , the adversary (Eve)
waits until he has seen the ciphertexts $\left\{(\mathbf{s}_1,\mathbf{t}_1), (\mathbf{s}_2,\mathbf{t}_2), \cdots, (\mathbf{s}_{i-1},\mathbf{t}_{i-1})\right\} $
(which he lets pass unchanged to the receiver) and then creates and sends a fraudulent ciphertext $(\mathbf{s}_i,\mathbf{t}_i)$ which he
hopes to be accepted by the receiver as the $i$th ciphertext.

Essentially, Eve's strategy is to maximize the false acceptance rate by selecting a suitable message $\mathbf{s}_i$ and a key $\mathbf{k}_E$, namely,
\begin{eqnarray}
  \label{eq:cn}
     \max_{\mathbf{s}_i \in \mathcal{S},\mathbf{k}_E\in \mathcal{K}} E\{\eta|H_1,\mathbf{s}_i,\mathbf{k}_E\}.
\end{eqnarray}
Equivalently, this means
 \begin{eqnarray}
  \label{eq:cn}
     \min_{\mathbf{s}_i \in \mathcal{S},\mathbf{k}_E \in \mathcal{K}} d_H\left(\tau(\mathbf{s}_i,\mathbf{k}_B),\tau(\mathbf{s}_i,\mathbf{k}_E)\right),
\end{eqnarray}
as shown by (\ref{eq:etaH1}).

\begin{lem}
In order to minimize the false acceptance rate of an impersonate attacker, the minimum Hamming distance of the code ensemble $\Omega(\mathcal{C})$, namely,
$d_{\min}\left(\Omega(\mathcal{C})\right)$, should be maximized.
\end{lem}

\section{Numerical Example}
Consider a physical-layer authentication system, in which binary key is of length $L_k=128$ and the authentication tag is of length $L_t=256$. To attack this physical-layer authentication system, a potential eavesdropper tries to do her or his best to decode the key. As one can consider the block codes of rate $R_c$ over a Bi-AWGN channel of the SNR $\gamma_t$, the equivalent $E_b/N_0$ can be defined as $E_b/N_0 = R_c^{-1} \gamma_t$.

\begin{figure}[htb] %[htbp]
   \centering
   \includegraphics[width=0.5\textwidth]{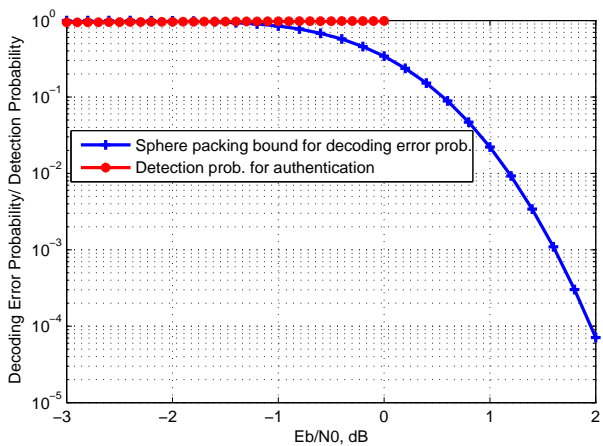} %{PWBF_BER_Uth.eps}
   \caption{Sphere-packing low bound on the decoding error probability and detection probability (or successful authentication rate) versus $E_b/N_0$.}
   \label{fig:spb}
\end{figure}

Fig. \ref{fig:spb} shows the SP59 bound on the decoding error probability and detection probability (successful authentication rate) for different $E_b/N_0$'s.  As the eavesdropper cannot do better than a ML decoder, the SPB bound provides an over-estimate of its capability on guessing the key. As shown, the eavesdropper becomes hopeless in guessing the key whenever $E_b/N_0$ is below to about -1 dB as the decoding error probability is around 1. However, the authentication system does work well with almost perfect successful authentication rate.  In simulations, the threshold is set so as the false alarm probability is lower than 0.01.

\section{Conclusion}

We propose a channel coding approach for physical layer authentication. With this new approach, the computational security for classic authentication schemes can be well formulated using a new decoding approach. The well-designed physical-layer authentication can ensure a new degree of security, namely, information-security, thanks to the introduction of channel noises during transmission.

For design of a physical layer authentication system, the success authentication rate should be balanced with an acceptable false acceptance rate. It is beneficial for use of long tags, as the success authentication rate can be enhanced while the false acceptance rate can be reduced. In the meantime, numerical results show that even with short tags (of length 256), the best possible decoding error probability under ML decoding can approach 1 while the authentication still work well.

\section*{Acknowledgment}
This work was supported in part by the National Natural Science Foundation of China under Grants 61372123, 61271335, 61032004, 61302103. The work of Wu was also supported by the Scientific Research Foundation of Nanjing University of Posts and Telecommunications under Grant NY213002. The work of Yang was also supported by the Key University Science Research Project of Jiangsu Province under Grant 14KJA510003.

%\bibliography{IEEEabrv,bib_wu}

\end{document}